# Towards Approximate Mobile Computing

Veljko Pejović,
Faculty of Computer and Information Science,
University of Ljubljana, Slovenia

When Dennard scaling, a law describing the area-proportional growth of integrated circuit power use, broke down sometime in the last decade, we faced a situation where further transistor minimization suddenly required additional energy for operation and cooling. CPU manufacturers responded with multicore processors, as an alternative means to increase the floating-point operations per second (FLOPS) count. However, this too increases the energy consumption and, in addition, requires a larger silicon area. The most threatened by the stalled growth of per-Watt computing performance are pervasive mobile computers, nowadays present in anything from wearables to smartphones. Not only does these devices' small form factor prevent further component packing, but the need for mobility also precludes bundling devices with large batteries.

Yet, a unique opportunity for optimization lurks behind the *mobile* aspect of today's computing. With computation executed in an array of environments, **user expectations with respect to result accuracy vary, as the result is further manipulated, interpreted, and acted upon in different contexts of use**. For instance, a user might tolerate a lower video decoding quality when calling to say "Hi" from a backpacking holiday, while she would expect a higher video quality when on a job interview call from an office. Similarly, when searching for nearby restaurant suggestions, rough location determination and a slightly shuffled ordering within the produced suggestion list would probably go unnoticed, whereas the same inaccuracies would not be tolerated when driving directions are searched for.

**The result of a computation need not be perfect, just good enough for things to work.** This opens up opportunities to save resources, including CPU cycles and memory accesses, thus, consequently battery charge, by reducing the amount of computation to the point where the result accuracy is just above the minimum necessary to satisfy a user's requirements. This way of reasoning about computation is termed *approximate computing (AC)* and *Approximate Computing Techniques (ACTs)* have already been demonstrated on various levels of computer architecture, from the hardware where incorrect adders have been designed to sacrifice result correctness for reduced energy consumption [1], to compiler-level optimizations that omit certain lines of code to speed up video encoding [2]. Experiments have shown significant resource savings, e.g. tripled energy efficiency with neural network-based approximations [3], or 2.5 times the speedup when certain task patterns are substituted with approximate code [4]. Ironically, to date, approximate computing remains mostly confined to desktop and data center computing, missing the opportunity to bring the benefits to mobile computing. It is exactly in this domain where, due to context-dependent user requirements the occasions for adaptable approximation are abundant, and where, due to the devices' physical constraints, the applicability of alternative solutions for increasing the computational capacities, such as further component packing, is the lowest.

Recently, the necessary conditions for the emergence of a new paradigm – *Approximate Mobile Computing (AMC)* – have been all but fulfilled. First, hardware capabilities of mobile devices have reached the level that allows very complex on-device computation. This is especially true in the area of artificial intelligence, where neural processing units (NPUs), such as Qualcomm Zeroth, allow deep learning algorithms to be run locally on the device. Second, the growing popularity of mobile personal assistant applications, e.g. Google Assistant, Siri, Cortana, and Amazon Alexa, opens up opportunities for inexact computation. These apps are tightly integrated with the user, operate in varying contexts, are used for queries where no golden answer exists (e.g. for content suggestions), and rely on inherently probabilistic natural language processing and computer vision algorithms. Finally, as we turn to our mobile devices for a wider range of tasks, over longer periods of time, and in increasingly diverse situations, we are in a position to better understand users' expectations from mobile computation.

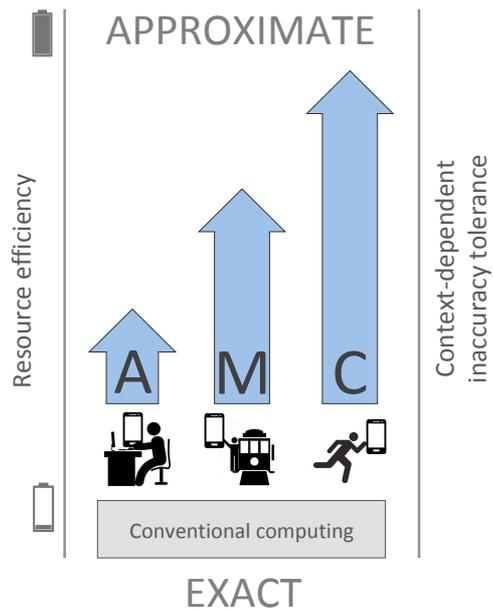

Figure 1: Approximate mobile computing (AMC) departs from the rigidness of conventional computing and increases resource efficiency by enabling a controlled reduction in result accuracy to the point defined by a user's context-dependent inaccuracy tolerance.

Making AMC a reality requires that we first resolve key doubts about how to enable approximation on mobile devices, how to infer a user's context-dependent result accuracy expectations, and how to adjust the approximation so that the expectations are met in the most resource-efficient way. In the rest of the article we analyze the state of the art along these fronts and derive guidelines for future efforts in each of the fields.

## STARTING POINT – CONVENTIONAL APPROXIMATE COMPUTING

A range of approximate computing techniques (ACTs) operating at all levels of the computing stack have been developed in the last ten years [5],. *Hardware layer techniques* include, for instance, approximate circuits, such as adders and multipliers that use low-precision transistors for operations on the least important bits, and thus reduce energy requirements while sacrificing only a limited amount of result accuracy [1]. A technique presented in [6] is based on the observation that changes in high-order bits of video data tend to be easier to detect by the human eye than changes in low-order bits of data. High-order bits of pixel data are thus stored in reliable memory segments, while low-order bits go to less reliable memory (Figure 2). The difference between the segments is in the DRAM refresh rate – the higher the rate, the more reliable the segment is, but more energy is needed for the storage.

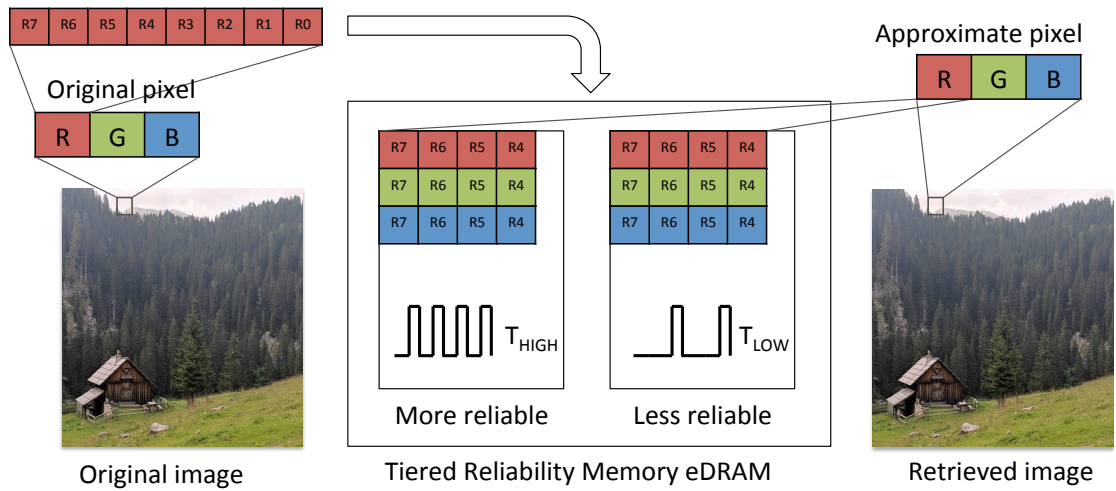

Figure 2: Tiered reliability memory saves energy by storing low-order bits of pixel data into less reliable lower refresh rate ($T_{LOW}$) memory segments (adapted from [6]).

One of the earliest *software layer techniques* has been proposed by Rinard [7]. Here, a program is written as a set of tasks, whose execution can be discarded, should this lead to the execution speedup without a significant impact on the result quality. Another technique, approximate memoization [8], stores a limited number of function execution results, so that for subsequent function executions with similar input parameters one of the pre-calculated results is returned. The approximation can also be moved further down the stack. Loop perforation [2], a method developed at MIT that skips some of the loop's iteration in order to reduce the amount of computation and save resources, has been implemented at the compiler level, enabling automatic application of the technique on selected loops.

## CAN WE IMPLEMENT APPROXIMATE COMPUTATION ON MOBILES?

The applicability of the above techniques to mobile devices must be examined through the lens of mobile computing constraints. Smartphones are highly versatile and expected to run an array of different applications in parallel. Many, especially hardware-based, techniques are often not flexible enough to support a mix of concurrently executed applications. A smartphone user might tolerate imperfect rendering in a 3D game, but data encryption protocols require perfectly accurate computations. One solution is to fit devices with both exact and approximate versions of the same hardware. However, this clashes with the portability-driven need for maintaining a small form factor. Another constraint comes from mobile apps' interactivity – an average session with a smartphone lasts between 10 and 250 seconds, while an average user performs 10 to 200 such sessions in a day. This restricts applicable ACTs to those that are quick to set up and trigger.

**Challenges:** the main obstacles towards exploring the benefits of approximation on mobiles are the lack of ACT implementations for mobile systems and the lack of support for writing and building approximate programs on mobile platforms. Regarding the former, selected ACTs need to be implemented in general frameworks for mobile application development. This could include the implementation of loop perforation at the level of the LLVM compiler used for compiling iOS applications, or supporting GPU processing kernel substitution with approximate implementations in NVIDIA CodeWorks for Android. Regarding the software writing support, ACTs often expect a developer to explicitly define parts of the program that may be executed approximately. Frameworks, such as Green [9], allow a developer to use C++ annotations to both provide approximate versions of the code (e.g. alternative function implementations), as well as to indicated approximable code blocks (e.g. loops that need not be executed with a full number of iterations). The annotations are then used to instruct the compiler to generate a suitable approximate version of the program.

## CAN WE TELL IF A USER IS SATISFIED WITH THE RESULT QUALITY?

Opportunities for approximation arise only when a user is satisfied with sub-accurate computation results. For instance, a user expects an activity tracking wristband to accurately monitor vital signs and recognize different movement patterns while exercising, yet the battery charge can be saved during non-exercise times, when the user merely expects the wristband to recognize a step so that the total count is taken.

Pervasive use of mobile computing allows us to inspect how a user's satisfaction with the delivered computation result changes with the context of use. Numerous aspects of the situation and the environment can impact a user's perception of the result. Thus, we consider "context" to be a complex term, a view of which we obtain through a mobile device's built-in sensors. For example, we can sense a user's physical activity via a phone's accelerometer, location via GPS, through a combination of light and location sensors we can infer whether a user is indoor or outdoor, and so on. Coordinating frequent sampling of a multitude of a device's sensors, and storing and transferring the data can be a tedious task, with which dedicated sensing frameworks, such as AWARE [10] can help.

In the second step, we need to monitor the use of an AMC application, deliver results of varying quality, and obtain information about a user's satisfaction with the delivered result. Mobile experience sampling method (mESM) allows us to query the user about her recent experience immediately after the app usage session [11]. A well-designed study can minimize the number of queries and ask the user about the experiences in previously unseen situations only. The exact flavor of the questions asked depends on the application. Voice/video communication applications (e.g. Skype, Whatsapp), for example, use simple Likert-scale questions (e.g. number of stars corresponding to the quality of the call) to get a quick feedback on the service quality.

Finally, machine learning lets us establish the link between the context, sensed at the time of querying, and the mESM answers, in order to model the change in result quality expectations in different situations. Such a model could, for example, learn that a user is satisfied with the personal assistant's voice command comprehension, even if the speech recognition was run on an approximate neural network, as long as the app is used at home in the evening (we hypothesize that the lack of noise in the environment and a limited, predictable set of queries a user might issue in such a situation, e.g. "Set alarm for 8am", could be a confounding factor for a user's satisfaction).

**Challenges:** Context sensing is one of the most energy expensive operations on a mobile phone. To capture user's expectations in different situations, sensing and mESM querying might have to be performed each time the app is used. Furthermore, numerous aspects of the context can impact a user's reaction, thus, sensing needs to be comprehensive and involve as many sensors as possible. For instance, a video call decoding quality requirements might depend on the level of outdoor brightness, the mode of transport that a user is on, but also on the relationship with the other party, or even the nature of a conversation. A potentially very large space defined by relevant contextual variables represents a major challenge, since frequent modification of the result accuracy, followed by querying, might irritate the user. Techniques such as active learning, where a user is queried about her experiences only if the existing model is unsure about the user's reaction, or reinforcement learning that controls both the accuracy adaptation and querying so to optimize a reward related to a user's satisfaction and resource use represent interesting research avenues.

## CAN WE DYNAMICALLY ADAPT AMC TO MAXIMISE RESOURCE SAVINGS WHILE STILL SATISFYING A USER'S RESULT QUALITY EXPECTATIONS?

As discussed above, approximation may be tolerated only in certain situations. Consequently, we need a means for dynamic adaptation of the result precision. Such adaptation "knobs" have already been implemented with certain ACTs. Hoffmann et al. "hijack" and expose a for-loop iteration counter increments, so that a variable number of loop iterations can be skipped [13]. More effort is needed to expose similar "knobs" for numerous other ACTs.

Once the knobs are exposed, we must know how to set them to achieve the desired result quality, as different amounts of approximation lead to different result accuracy and resource savings. Misailovic et al. built a Quality of Service (QoS) profiler that for a given program, given test input, and a QoS metric calculates the loss of accuracy and the overall speedup under different approximation levels (brought by loop perforation) [2]. Combined with the model that describes how a user's expectations depend on the context, the profiler output tells us how to set the approximation knobs in order to achieve the maximal savings and ensure that the result is acceptable for the user.

**Challenges:** Despite prior accuracy profiling, approximation adaptation needs to be recalibrated according to the run-time performance. Due to the discrepancies between the test and the actual input data, or due to a potential impact of the context on the calculation, the calculated result quality might not reach the previously estimated levels. However, even assessing the result quality is often expensive. In most situations, we can evaluate the quality only if the result of a perfectly accurate computation is available, defying the purpose of approximation. Laurenzao et al show that in image approximation it suffices to evaluate the result quality on small representative snippets of data [14], yet, this might not generalize to other domains. In addition, the app needs to have the information about the current context in order to adapt to it. The key question of AMC – whether the benefits enabled by approximate execution surpass the cost of context sensing and the adaptation – will be answered once the first AMC prototypes are completed and tested.

## THE ROAD AHEAD

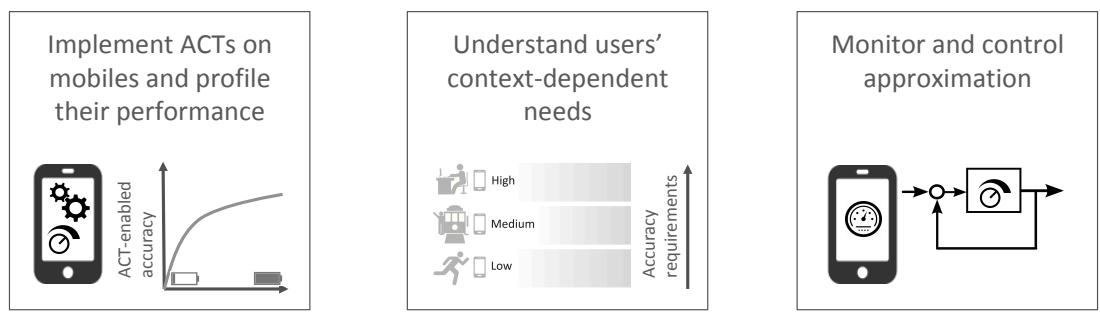

Figure 3: Key steps towards Approximate Mobile Computing.

The overarching goal of AMC is to enable further proliferation of mobile computing by drastically reducing the resource requirements of modern apps, while ensuring that users' needs are satisfied. Opportunities for approximation appear in certain contexts, as they are conditioned on a user's context-dependent perception of the result. In Figure 3 we sketch three broader directions that research should take in order to make AMC the reality. Namely, 1) implementing ACT support and understanding the benefits of approximation in mobile devices, 2) building a framework for sensing the context, querying the user's expectations, in order to model the relationship between the context and users' accuracy needs, and 3) devising a system for monitoring and controlling the approximation.

In this article we raised certain challenges pertaining to each of the steps. However, they are by no means exhaustive, nor detailed enough. For instance, resource savings brought by a single app's modification are notoriously difficult to evaluate on mobile devices, as the cost of a component usage (e.g. a GPS chipset) depends on its previous state, which may be affected by other apps on the phone [15].

Yet, the main challenge of AMC stems from its highly interdisciplinary nature. Efforts by computer architecture, compilers, and programming languages experts are needed to bring ACTs to mobiles; human-computer interaction (HCI) and mobile sensing experts can help with understanding users' result accuracy expectations; mobile system and control theory experts should contribute towards controlling dynamic approximation adaptation. The topic of approximate computing has already gained a lot of traction within programming languages, formal verification, and computer architecture communities. This is witnessed by a number of specialized workshops, such as "Workshop On Approximate Computing" with "High Performance and Embedded Architecture and Compilation Conference (HiPEAC)", and "Workshop on Approximate Computing Across the Stack" with "Programming Language Design and Implementation Conference (PLDI)", as well as special journal issues on the topic, such as a recent IEEE Micro Approximate Computing issue. However, to date, mobile computing, mobile sensing, and mobile HCI communities left the topic virtually untouched. With this article we hope to start the conversation and mobilize a wider research community towards making approximate mobile computing a reality.